\documentclass[manuscript]{aastex}

\include{amssymb}

\shorttitle{Dead Zones as Thermal Barriers to Rapid Planetary Migration}
\shortauthors{Hasegawa \& Pudritz}

\begin{document}

\title{Dead Zones as Thermal Barriers to Rapid Planetary Migration in Protoplanetary Disks}

\author{Yasuhiro Hasegawa and Ralph E. Pudritz\altaffilmark{1}}
\affil{Department of Physics and Astronomy, McMaster University,
    Hamilton, ON L8S 4M1, Canada}
\email{YH:hasegay@physics.mcmaster.ca, REP:pudritz@physics.mcmaster.ca}

\altaffiltext{1}{Origins Institute, McMaster University, Hamilton, ON L8S 4M1, Canada}

\begin{abstract}
Planetary migration in standard models of gaseous protoplanetary disks is known to be very rapid 
($\sim 10^5$ years) jeopardizing the existence of planetary systems. We present a new mechanism 
for significantly slowing rapid planetary migration, discovered by means of radiative transfer calculations 
of the 
thermal structure of protoplanetary disks irradiated by their central stars. Rapid dust settling 
in a disk's dead zone - a region with very little turbulence - leaves a dusty wall at its outer edge. 
We show that the back-heating of the dead zone by this irradiated wall produces a positive gradient of 
the disk temperature which acts as a thermal barrier to planetary migration which persists for the disk 
lifetime. Although we analyze in detail the migration of a Super-Earth in a low mass disk around an M star, 
our findings can apply to wide variety of young planetary systems. We compare our findings with other 
potentially important stopping mechanisms and show that there are large parameter spaces for which dead 
zones are likely to play the most important role for reproducing the observed mass-period relation in longer 
planetary periods. 
\end{abstract}


\section{Introduction}

Extrasolar planets (ESPs) have an unexpected distribution of orbital radii around their host 
stars \citep{ufq07} - ranging from about 0.02 to 70 astronomical units 
(AU).\footnotemark \footnotetext{See the website: http://exoplanet.eu.} 
In particular, ESPs are observed to obey a mass-period (M-P) relation wherein lower mass planets
end up in short period orbits around their host stars \citep{us07}. The predominance of very short period 
planets is generally thought to arise as a consequence of planetary migration.  As an example, the 
tidal interaction of a planet with its surrounding gaseous disk excites density waves in the disk at 
so-called Lindblad resonances. These waves exert a torque back on the planet which results in a net angular 
momentum transfer between them \citep{gt80}. Planets may also exchange angular momentum with the gas 
inside of their horseshoe region \citep{ward91}. For locally isothermal protoplanetary disks with 
smoothly declining distributions of disk column density and temperature with radius, 
the net torque generally leaves a planet spiraling inwards through the disk \citep{ttw02}, 
i.e. the torque exerted by the outer wake is marginally stronger than that of the inner wake \citep{ward97}.  
Many calculations and simulations show that the migration timescale of planets in such "standard" disk 
models is very short - roughly two orders of magnitude smaller than the disk lifetime (one to ten 
million years (Myr)) \citep{ward97,npmk00,ttw02,dkh03}. Why are there any planetary systems at all? 
 
The key to understanding the M-P relation and the survival of planetary systems is in how the dynamics of 
planetary motion is coupled to the properties and structure of the protoplanetary disks. As an example, 
\citet{sli09} focused on the surface density transition that can be produced at the location of the ice-line, 
where a local pressure maximum can act as an accumulation point for planetesimals \citep{kl07}. If it is 
assumed that type I migration is  much slower than predicted in locally isothermal disk models, this 
feature could account for planets with orbital radii 0.1 - 2 AU. Obviously, a physical explanation for 
slower migration is needed.

In this Letter, we present a new slowing mechanism of rapid type I migration -which may occur for 
planets with masses that are too low to open up a gap in their disks (massive planets can tidally 
form a gap and undergo type II migration). We show, by means of Monte Carlo radiative transfer simulations, 
that dead zones - the dense inner disk region wherein turbulence cannot be readily excited \citep{g96} - 
support a thermal barrier to migration. One of the most important consequences is that the thermal barrier 
could account for planets at larger orbital radii. In $\S$ 2, we outline our disk model 
and discuss tidal torques. In $\S$ 3, we analyze how the presence of a thermal barrier impacts the migration 
rates of low-mass planets. In $\S$ 4, we discuss potential issues for the M-P relation by comparing our 
stopping mechanism with others.

\section{Disk model \& tidal torques}

Protoplanetary disks are known to be heated by radiation from the central star 
\citep[hereafter, HP10]{cg97,dccl98,hp09b}. This radiation mainly 
determines the thermal structure of disks \citep{kh87}. This is because viscous heating dominates stellar 
irradiation heating only within about 1 AU for the classical T Tauri star systems \citep[CTTSs;][]{dccl98} 
and only within about 0.1 AU  for lower mass stars such as M stars (HP10).

Detailed modeling of the spectral energy distributions emitted by disks has shown that $s=-1$ for 
disks where the disk surface density $\Sigma \propto r^{s}$ \citep{dccl98}. It is well 
established that the sign of a net torque exerted on planets depends on 
two central properties of disks, their surface density and temperature \citep[e.g.,][]{ward97}. 
The thermal structure therefore plays a critical role in controlling the direction of planetary migration. 
In disks without internal structure, the disk temperature $T \propto r^{t}$ at the mid-plane steadily 
decreases ($t < 0$) and planetary migration is steadily inwards.  

The point is that disks are not simple power-law structures. The strength of turbulence within them 
varies considerably, with very low levels occurring in dense regions called dead zones \citep{g96} that 
initially extend over roughly 10 AU in disks \citep{mp06} and then gradually shrink in size as disk 
material is accreted onto the central star \citep[hereafter, MPT09]{mpt09}.  
Turbulence in disks is most likely excited 
by the magnetorotational instability (MRI) \citep{bh91b}. The MRI requires good coupling between ions and 
magnetic fields, which is largely absent in the dead zone - that inner, high density region in which the 
ionization due to the X-rays from the central star and cosmic rays is suppressed. 

Dust is the dominant 
absorber of stellar radiation in disks although its total mass is 100 times smaller than that of 
gas \citep{dhkd07}.  Many observations imply that it has a density distribution that is different from 
the gas distribution, which is derived assuming vertical hydrostatic equilibrium \citep{kh87}.  
Dust settling, a consequence of its size distribution \citep{dd04b}, is ubiquitous in protoplanetary 
disks around any young star (HP10, references herein). The dust scale height depends upon the amplitude 
of disk turbulence which keeps it suspended in the gravitational field of the disk (which is 
determined by the central star) \citep{dms95}.  

We demonstrated in HP10 that dust settling in dead zones results in the appearance of a limited 
region in which the disk temperature can actually increase with radius - a radial temperature inversion. 
In this Letter, we adopt the disk model developed in HP10 (see Table. \ref{param_disk}) and 
focus our computations on M dwarf systems, such as the recently discovered Super-Earth 
($\sim 5 M_{\oplus }$; $M_{\oplus }$ is the Earth's mass) with an orbital radius of 2 AU \citep{bbfw06}.  
We refer the readers to HP10 for the detail. The low mass of disks around M stars allows 
much more comprehensive Monte Carlo simulations to be performed, but our analysis in principle applies
to any protoplanetary disk.

We adopt the torque formula \citep{ward97,mg04,js05} 
in which only the Lindblad torque is considered. This is because the corotation torque is readily 
saturated  (i.e. is canceled out)
in dead zones in our radiatively heated disk model. We compare the libration 
timescale (ie the timescale for gas to complete an orbit in the horseshoe region), 
$\tau_{lib}\approx 8\pi r_p/3\Omega_p x_s$, with the viscous timescale, 
$\tau_{vis}\approx x_s^2/3\nu$, where the half-width of the horseshoe region is 
$x_s/r_p=1.68(M_p r_p/M_* h_p)^{1/2}$ \citep{pp09}, the kinematic 
viscosity is $\nu=\alpha h^2 \Omega_{Kep}$ \citep{ss73}, and the disk scale height is $h$. 
Our numerical results give $h_p/r_p\simeq 0.05$, so that the critical value of turbulence, $\alpha_{crit}$, 
below which the corotation torque is saturated (and therefore negligible), is
\begin{equation}
 \alpha_{crit}=0.01 \left( \frac{M_p}{5M_{\oplus}} \right)^{3/2} \left( \frac{M_*}{0.1M_{\odot}} \right)^{-3/2}
                    \left( \frac{h_p/r_p}{0.05} \right)^{-7/2}.
\end{equation}
Since the dead zone has $\alpha=10^{-5}$, we can safely neglect the corotation torque for 
$M_p \gtrsim 0.5 M_{\oplus}$ in the dead zone of our disk model. In addition, we confirmed that, 
in the active region where the corotation torque is generally unsaturated, both Lindblad and corotation 
torques result in inward migration in our disk model \citep{pbck09}. Thus, exclusion of the corotation 
torque in the active region does not affect our findings in our disk model. 
We note that corotation torque may be unsaturated in 
dead zones for sufficiently small planetary masses, but the exact limit will depends on knowing the disk 
scale height that is established by the host star. Furthermore, our torque formula takes into account 
the effects of vertical disk thickness by diluting the gravitational force of a planet by $z$ \citep{js05}.

\section{Results}

We performed numerical simulations of the thermal structure of 2D disks by solving the radiative transfer 
equation with a Monte Carlo method \citep[HP10]{dd04a}.  We included the effects of vertical hydrostatic 
balance, dust settling, a dead zone, and the gravitational field of a planet embedded in the disk.  
The tidal torque is calculated as described in $\S$ 2, and incorporates our numerical data, in order to 
calculate the migration time.

\begin{figure}[!ht]
\begin{center}
\includegraphics[width=10cm,height=5cm]{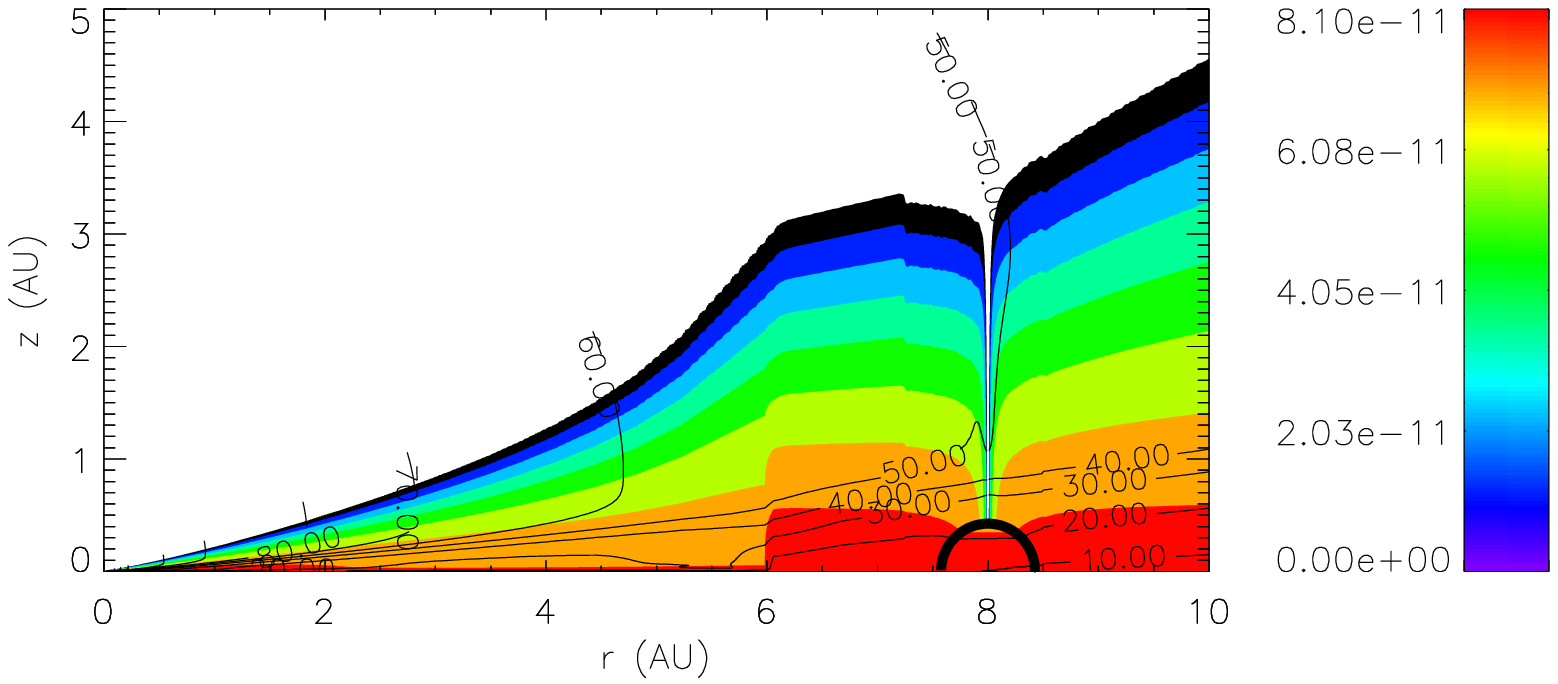}
\includegraphics[width=10cm,height=5cm]{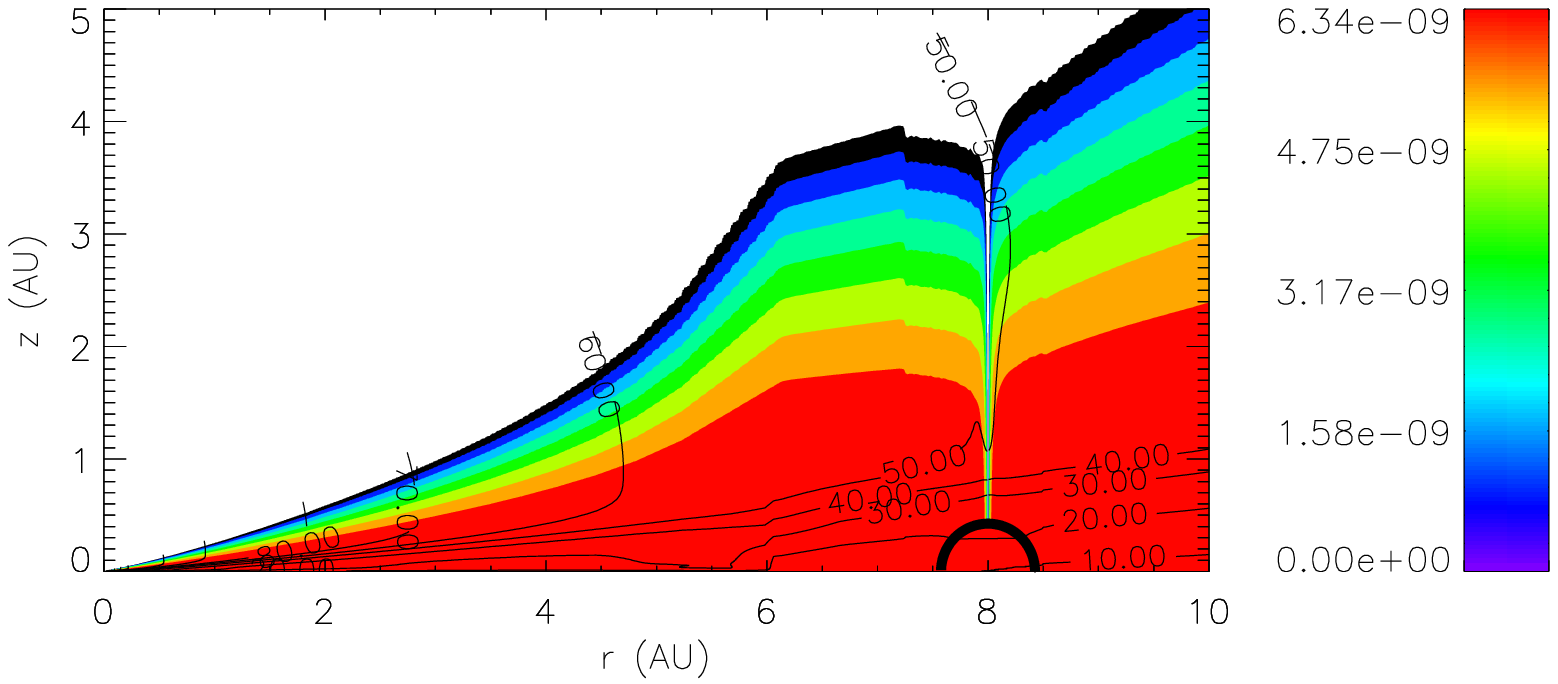}
\caption{The density and temperature structures of disks with a $5M_{\oplus }$ mass planet located 
at 8 AU. For both panels, the density is denoted by colors [g cm$^{-3}$] as shown in colorbar, and 
the disk temperature is denoted by contours [K] (both panels show the identical temperature since the disk 
temperature is defined by taking the mass-average of the dust temperatures). The size of a dead zone is 
6 AU. Top: the dust density. Bottom: the gas density. The thick black 
circle denotes the Hill sphere $r_H = r_p(M_p/3M_*)^{1/3}$, where $r_p$ and $M_p$ is the position and 
mass of a planet, respectively, and $M_*$ is the stellar mass. A wall-like structure appears 
at the boundary between the active and dead zones in the dust distribution while it is not in the gas.}
\label{fig1}
\end{center}
\end{figure}

\begin{figure}[!ht]
\begin{center}
\includegraphics[width=10cm,height=5cm]{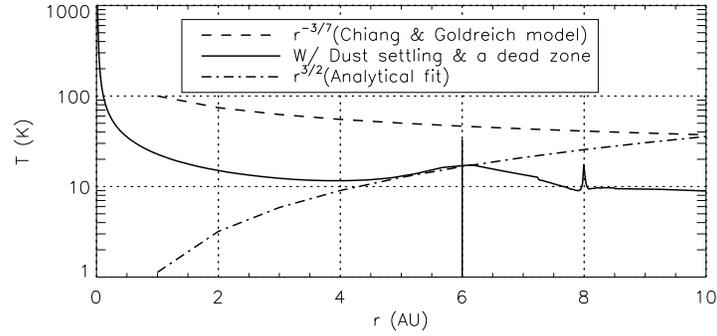}
\caption{The disk temperature in the disk mid-plane. The solid line is for the  case of dead zone, the 
dashed line is the analytical model of disk temperature \citep{cg97}, and the dashed-dot line is for 
analytical fit to a positive temperature gradient. 
The size of a dead zone, which is 6 AU, is indicated by the vertical solid line.}
\label{fig2}
\end{center}
\end{figure}

Figure \ref{fig1} presents the thermal and density structure of a disk with a $5 M_{\oplus }$ planet placed 
at 8 AU. The top and bottom panels show the dust and gas densities by color, respectively. Since we 
define the disk temperature as the mass-averaged temperature of dust, both panels show the identical 
temperature structure which is represented by contours. The thick line denotes the Hill radius 
$r_H = r_p(M_p/3M_*)^{1/3}$. In this Letter, we adopt, without loss of generality, 
a dead zone which is 6 AU in size. 
Dead zones enhance dust settling because of the low turbulent amplitude there.  Since disks have inner 
dead, and outer active regions for turbulence, the transition of the density distribution of dust occurs 
at the outer edge of the dead zone.  This leaves a marked step in the dust scale height behind - in 
effect a wall of dust.  The additional stellar energy absorbed at the wall is distributed by radiative 
diffusion  \citep{hp10} since the optical depth at this region is high. The resulting radial 
"thermal inversion" 
- ie a region of increasing temperature with increasing disk radius - is shown in Figure \ref{fig2}.  
An analytic fit to our data shows that this back-heated region in the dead zone has a positive 
temperature gradient described by a power-law $T\propto r^{t'}$ with $t'>3/2$. 

\begin{figure}[!ht]
\begin{center}
\includegraphics[width=10cm,height=10cm]{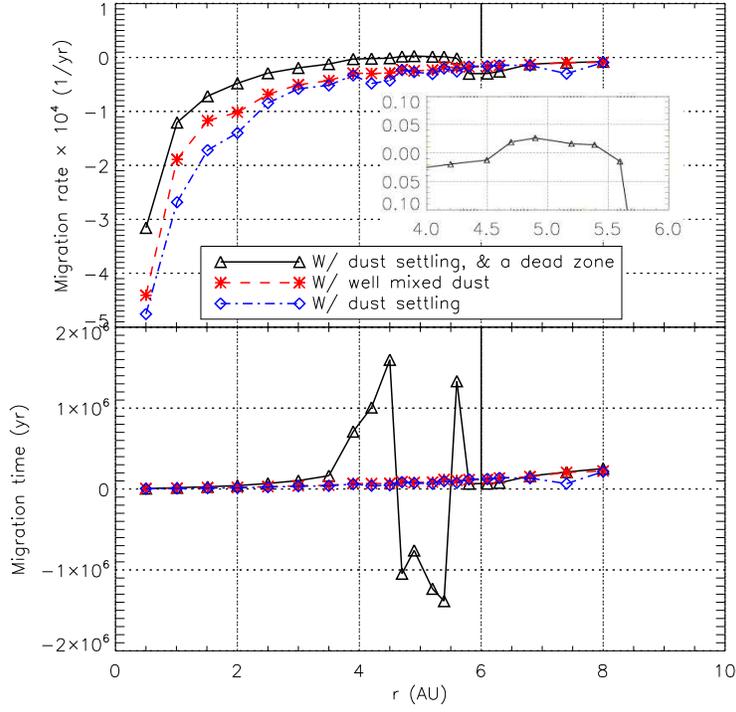}
\caption{Migration rate and time for a 5 mass $M_{\oplus}$ planet at various orbital radii on the top
and bottom panels, respectively. For both panels, the solid line denotes the 
case of a dead zone which is 6 AU in size whose position is indicated by the vertical solid line, the red 
dashed line is for the well-mixed case, and the blue dashed-dot line is for the dust settling case. The 
negative migration time region in the bottom panel (outward migration) corresponds to the region with 
a positive temperature gradient (see Figure \ref{fig2}).}
\label{fig3}
\end{center}
\end{figure}

We show in Figure \ref{fig3} (top) that this radial thermal inversion causes the migration rate to 
be positive (the migration time becomes negative - bottom panel), - i.e. planets migrate {\it outward} 
in the region with the positive temperature gradient. The physical explanation 
of this behavior is that the increasing function of disk temperature changes the disk's pressure 
distribution which in turn causes the position of the outer Lindblad resonances to be further from 
the planet than the inner ones \citep{a93}. This results in outer torques that are much weaker than the 
inner ones.\footnotemark 
\footnotetext{We will show in a forthcoming paper that the Lindblad torque becomes negative 
when $s-t/2<-7/4$ \citep{hp10}.} 
Figure \ref{fig3} (bottom) also shows that the planets very slowly enter the region of 
torque reversal - as seen by the strong positive "spike" in the migration time. These regions correspond 
to radii at which $dT/dr\simeq 0$ (see Figure \ref{fig2}) at the inner and outer resonances, making 
the torque difference between them very small.  

We emphasize that the positive temperature gradient arising from the wall-like dust structure 
is achieved for the case of a finite transition region, $\bigtriangleup r \leq 10 h$, in the value of 
turbulent $\alpha$ (HP10). Although, for simplicity, we adopt a sharp spatial transition from the active 
to the dead zone in this Letter ($\bigtriangleup r=0$), our above results are valid for the case of 
$\bigtriangleup r \simeq h$, since the positions of Lindblad resonances are typically offset from 
the planets by $2h/3$ \citep{a93}. In addition, it is interesting that the migration timescale of 
the M star system is similar to that of CTTS ($\sim 10^4-10^5$ years) for the other two cases 
(well mixed, dust settling). This is because the tidal torque is scaled by $\Sigma(h/r)^{-2}$. A more 
detailed discussion of them is presented in \citet{hp10}.

\section{Discussion}

\subsection{A thermal vs a density barrier at outer dead zone radius}

We find that the dusty wall produces a radial temperature inversion that is a thermal barrier for 
rapid type I planetary migration. 
Whereas the torque balance in the well coupled active zone forces planets to migrate inward, 
once they encounter the radial thermal inversion region, the torque balance reverses, and they move 
out of the region.  Thus, planets are trapped there if they originally migrate 
from the active region beyond the dead zones, or even if they formed close to the outer edge of 
the dead zone.

The astrophysical implications of this result are very important since we have shown that dusty 
protoplanetary disks with dead zones possess an innate mechanism for strongly slowing planetary 
migration within them, provided that the corotation torque is saturated. 
While such a thermal barrier exists for any size of dead zone (HP10), 
its effectiveness is probably most important for lower mass disks, as we now show. 

The density structure of disks evolves with time due to viscous evolution. MPT09 found that the 
difference of $\alpha$ between the active and dead regions produces a steep density gradient at 
the boundary and the location of their jump moves inward with time over the long ($\sim$ 10 Myr) viscous 
timescale of the dead zone. This density gradient 
at the outer dead zone boundary also plays an important role in slowing down or stopping type I migration, 
provided that planets migrate from larger disk radii. The inner torques become larger than the outer ones 
in the density gradient region, resulting in the reflection of migrating planets off the density 
gradient (MPT09).

The relative importance of these two dead zone mechanisms is controlled by the ratio of dust settling 
$\tau_{set}\approx \Sigma / \sqrt{2\pi} \rho_d a \Omega_{Kep}$ and the viscous $\tau_{vis}$ 
timescales,  
where $\rho_d$ is the bulk density of dust and $a$ is the grain size of dust. We find that the critical 
condition $\tau_{set} / \tau_{vis} \geq 1$ for dominance of the density vs thermal barriers presented 
by a dead zone is
\begin{equation}
  \Sigma \left(\frac{h}{r}\right)^2  \geq   25 \left( \frac{\alpha}{10^{-2}} \right)^{-1}                                      
                                     \left( \frac{\rho_d}{1 \mbox{g cm}^{-3}}\right)
                                     \left( \frac{a}{1 \mbox{mm}} \right)\equiv C_{crit}.
\end{equation}
This implies that a density barrier is dominant for sufficiently large values of $\Sigma$ and $h/r$.
For disks around CTTSs with a typical dead zone size ($\approx$ 10 AU), 
$\Sigma (h/r)^2 \sim 300$ g cm$^{-2}\times (0.4)^2 \approx 2C_{crit}$ \citep{cg97},  
which indicates that a density barrier is dominant. For disks around M stars 
with a typical dead zone size ($\approx$ 5 AU), $\Sigma (h/r)^2 \sim 20$ g cm$^{-2}\times (0.05)^2 =2
\times10^{-3}C_{crit}$ \citep{sjw07}, which implies that a thermal barrier is dominant. Generally, we find 
that a thermal barrier becomes weaker in the late stages because of accretion which reduces 
the density at the outer edge of the dead zones.

\subsection{Comparisons with other possible stopping mechanisms and potential effects on the M-P relation}

It is well known that tidal interaction and angular momentum exchange with the central star \citep{lbr96} 
and the creation of a hole in the inner part of disks by the presence of a stellar magnetosphere 
\citep{sno94,lbr96} cannot reproduce the whole of the observed M-P relation. This is because 
such barriers become important only for planets approaching very close to the central star. Most ESPs, 
however, have observed orbital radii from about 0.02 AU to 10 AU \citep{us07}. Thus, while these barriers may 
play a role in piling up ESPs in the vicinity of the star, they have difficulty in 
predicting planets with larger orbital radii.

Stochastic migration provides another mechanism for controlling planetary migration. It arises 
when disks undergo magnetohydrodynamic turbulence \citep{np04,lsa04}, which is the outcome of 
the MRI \citep{bh91b}. Stochastic torques tend to reduce the timescale 
for planetary survival, however \citep{jgm06}. In a few exceptional cases, planets can diffuse out to 
large disk radii where they can survive longer. Planets within the dead zones cannot perform random 
walks because turbulent torques are reduced by about two orders of magnitude there and consequently 
would undergo steady inward type I migration \citep{omm07}. Thus, stochastic migration is unlikely 
to be the main barrier to rapid type I migration. 

Is our thermal barrier sufficiently wide to stop planets scattered into the dead zone by stochastic 
effects? We consider a characteristic length scale for turbulent diffusion defined by 
$\bigtriangleup r_{turb}=\sqrt{\nu \tau_c}=h\sqrt{\alpha \tau_c/\Omega^{-1}}$, where $\tau_c$ is 
the correlation time of turbulence. We adopt a 
value $\tau_c=0.5 \Omega^{-1}$ \citep{np04} and find that $\bigtriangleup r_{turb}\approx 0.02$ AU 
at $r=6$ AU. This is shorter than the width of the thermal barrier ($\sim 2$ AU). Therefore, stochastic 
motions due to turbulence are unlikely to affect the migration stopping mechanism at the thermal barrier. 

Corotation torques are also potentially important for slowing or stopping planets. In both barotropic and 
adiabatic disks, corotation torque associated with a radial vortensity gradient may work 
as a barrier around the 
region of inner edge of the dead zone \citep[$\sim$ 0.01-0.1 AU;][MPT09]{ftb02,mmcf06} because it 
becomes positive due to a positive surface density gradient, resulting in outward planetary migration 
\citep{mmcf06}. However, the location of the barrier is almost constant with time because 
stellar irradiation controls it, and consequently this barrier 
is important only for planets in the vicinity of the star. 
In adiabatic disks, corotation torque associated with a radial entropy gradient  may
also act as a barrier because it becomes large and positive around the region with a large (in magnitude), 
negative entropy gradient \citep{pm06,bm08,pm08,pp08}. Thus, corotation torque 
may be important to the M-P relation in certain regimes of planetary mass and disk heating, as noted above.

Our mechanism has a movable barrier that shrinks from larger disk radii on the 10 Myr (for CTTSs) viscous 
timescale of the dead zone. As a concrete example, this shrinkage of the dead zone could explain 
the recent detected Super-Earth at 2 AU around an M star \citep{bbfw06}. 
This is because a Super-Earth is likely to be the most massive planet that would surely form in a low 
mass disk and is therefore most likely to have been formed beyond the outer dead zone radius and 
left behind as the dead zone shrinks away. 

We conclude that the stopping mechanisms arising from dead zones - via thermal and density gradients - 
are important barriers to rapid planetary migration. We have shown that the thermal barrier 
that arises from disk irradiation by a heated dust wall is robust and may be most important in the earlier 
phases of disk evolution and for the evolution of low mass systems as found around M stars as an example. 
Dead zones may provide the most significant slowing mechanism of type I migration that is needed to explain 
the longer period of population of the M-P relation, which will be checked in future population 
synthesis models.

\acknowledgments
The authors thank Kees Dullemond, Thomas Henning, Hubert Klahr, Kristen Menou, Soko Matsumura 
and Richard Nelson for stimulating discussions.  We also thank an anonymous referee for a useful report. 
Our simulations were carried out on computer clusters of the 
SHARCNET HPC Consortium at McMaster University. YH is supported by McMaster University, as well as by 
Graduate Fellowships from SHARCNET and the Canadian Astrobiology Training Program (CATP). REP is supported by 
Discovery Grant from NSERC.

\bibliographystyle{apj}


\begin{thebibliography}{43}
\expandafter\ifx\csname natexlab\endcsname\relax\def\natexlab#1{#1}\fi

\bibitem[{Artymowicz(1993)}]{a93}
Artymowicz, P. 1993, \apj, 419, 155

\bibitem[{Balbus \& Hawley(1991)}]{bh91b}
Balbus, S.~A. \& Hawley, J.~F. 1991, \apj, 376, 223

\bibitem[{Baruteau \& Masset(2008)}]{bm08}
Baruteau, C. \& Masset, F.~S. 2008, \apj, 672, 1054

\bibitem[{Beaulieu {et~al.}(2006)}]{bbfw06}
Beaulieu, J.-P. {et~al.} 2006, \nat, 439, 437

\bibitem[{Chiang \& Goldreich(1997)}]{cg97}
Chiang, E.~I. \& Goldreich, P. 1997, \apj, 490, 368

\bibitem[{D{'}Alessio {et~al.}(1998)D{'}Alessio, Cant{\'o}, Calvet, \&
  Lizano}]{dccl98}
D{'}Alessio, P., Cant{\'o}, J., Calvet, N., \& Lizano, S. 1998, \apj, 500, 411

\bibitem[{D{'}Angelo {et~al.}(2003)D{'}Angelo, Kley, \& Henning}]{dkh03}
D{'}Angelo, G., Kley, W., \& Henning, T. 2003, \apj, 586, 540

\bibitem[{Dubrulle {et~al.}(1995)Dubrulle, Morfill, \& Sterzik}]{dms95}
Dubrulle, B., Morfill, G., \& Sterzik, M. 1995, Icarus, 114, 237

\bibitem[{Dullemond \& Dominik(2004{\natexlab{a}})}]{dd04b}
Dullemond, C.~P. \& Dominik, C. 2004{\natexlab{a}}, \aap, 421, 1075

\bibitem[{Dullemond \& Dominik(2004{\natexlab{b}})}]{dd04a}
---. 2004{\natexlab{b}}, \aap, 417, 159

\bibitem[{Dullemond {et~al.}(2009)Dullemond, Hollenbach, Kamp, \&
  D'Alessio}]{dhkd07}
Dullemond, C.~P., Hollenbach, D., Kamp, I., \& D'Alessio, P. 2009, Protostars
  and Planets V (Tucson: Univ. Arizona Press)

\bibitem[{Fromang {et~al.}(2002)Fromang, Terquem, \& Balbus}]{ftb02}
Fromang, S., Terquem, C., \& Balbus, S.~A. 2002, \mnras, 329, 18

\bibitem[{Gammie(1996)}]{g96}
Gammie, C.~F. 1996, \apj, 457, 355

\bibitem[{Goldreich \& Tremaine(1980)}]{gt80}
Goldreich, P. \& Tremaine, S. 1980, \apj, 241, 425

\bibitem[{Hasegawa \& Pudritz(2010{\natexlab{a}})}]{hp09b}
Hasegawa, Y. \& Pudritz, R.~E. 2010{\natexlab{a}}, \mnras, 401, 143

\bibitem[{Hasegawa \& Pudritz(2010{\natexlab{b}})}]{hp10}
---. 2010{\natexlab{b}}, in preparation

\bibitem[{{Jang-Condell} \& Sasselov(2005)}]{js05}
{Jang-Condell}, H. \& Sasselov, D.~D. 2005, \apj, 619, 1123

\bibitem[{Johnson {et~al.}(2006)Johnson, Goodman, \& Menou}]{jgm06}
Johnson, E.~T., Goodman, J., \& Menou, K. 2006, \apj, 647, 1413

\bibitem[{Kenyon \& Hartmann(1987)}]{kh87}
Kenyon, S.~J. \& Hartmann, L. 1987, \apj, 323, 714

\bibitem[{Kretke \& Lin(2007)}]{kl07}
Kretke, K.~A. \& Lin, D. N.~C. 2007, \apjl, 664, L55

\bibitem[{Laughlin {et~al.}(2004)Laughlin, Steinacker, \& Adams}]{lsa04}
Laughlin, G., Steinacker, A., \& Adams, F.~C. 2004, \apj, 608, 489

\bibitem[{Lin {et~al.}(1996)Lin, Bodenheimer, \& Richardson}]{lbr96}
Lin, D. N.~C., Bodenheimer, P., \& Richardson, D.~C. 1996, \nat, 380, 606

\bibitem[{Masset {et~al.}(2006)Masset, Morbidelli, Crida, \& Ferreira}]{mmcf06}
Masset, F.~S., Morbidelli, A., Crida, A., \& Ferreira, J. 2006, \apj, 642, 478

\bibitem[{Matsumura \& Pudritz(2006)}]{mp06}
Matsumura, S. \& Pudritz, R.~E. 2006, \mnras, 365, 572

\bibitem[{Matsumura {et~al.}(2009)Matsumura, Pudritz, \& Thommes}]{mpt09}
Matsumura, S., Pudritz, R.~E., \& Thommes, E.~W. 2009, \apj, 691, 1764

\bibitem[{Menou \& Goodman(2004)}]{mg04}
Menou, K. \& Goodman, J. 2004, \apj, 606, 520

\bibitem[{Nelson \& Papaloizou(2004)}]{np04}
Nelson, R.~P. \& Papaloizou, J. C.~B. 2004, \mnras, 350, 849

\bibitem[{Nelson {et~al.}(2000)Nelson, Papaloizou, Masset, \& Kley}]{npmk00}
Nelson, R.~P., Papaloizou, J. C.~B., Masset, F., \& Kley, W. 2000, \mnras, 318,
  18

\bibitem[{Oishi {et~al.}(2007)Oishi, {Mac Low}, \& Menou}]{omm07}
Oishi, J.~S., {Mac Low}, M.-M., \& Menou, K. 2007, \apj, 670, 805

\bibitem[{Paardekooper {et~al.}(2009)Paardekooper, Baruteau, Crida, \&
  Kley}]{pbck09}
Paardekooper, S.-J., Baruteau, C., Crida, A., \& Kley, W. 2009, preprint
  (astro-ph/arXiv:0909.4552v1)

\bibitem[{Paardekooper \& Mellema(2006)}]{pm06}
Paardekooper, S.-J. \& Mellema, G. 2006, \aap, 459, L17

\bibitem[{Paardekooper \& Mellema(2008)}]{pm08}
---. 2008, \aap, 478, 245

\bibitem[{Paardekooper \& Papaloizou(2008)}]{pp08}
Paardekooper, S.-J. \& Papaloizou, J. C.~B. 2008, \aap, 485, 877

\bibitem[{Paardekooper \& Papaloizou(2009)}]{pp09}
---. 2009, \mnras, 394, 2297

\bibitem[{Schlaufman {et~al.}(2009)Schlaufman, Lin, \& Ida}]{sli09}
Schlaufman, K., Lin, D. N.~C., \& Ida, S. 2009, \apj, 691, 1322

\bibitem[{Scholz {et~al.}(2007)Scholz, Jayawardhana, Wood, Meeus, Stelzer,
  Walker, \& O'Sullivan}]{sjw07}
Scholz, A., Jayawardhana, R., Wood, K., Meeus, G., Stelzer, B., Walker, C., \&
  O'Sullivan, M. 2007, \apj, 660, 1517

\bibitem[{Shakura \& Sunyaev(1973)}]{ss73}
Shakura, N.~I. \& Sunyaev, R.~A. 1973, \aap, 24, 337

\bibitem[{Shu {et~al.}(1994)Shu, Najita, Ostriker, Wilkin, Ruden, \&
  Lizano}]{sno94}
Shu, F., Najita, J., Ostriker, E., Wilkin, F., Ruden, S., \& Lizano, S. 1994,
  \apj, 429, 781

\bibitem[{Tanaka {et~al.}(2002)Tanaka, Takeuchi, \& Ward}]{ttw02}
Tanaka, H., Takeuchi, T., \& Ward, W.~R. 2002, \apj, 565, 1257

\bibitem[{Udry {et~al.}(2009)Udry, Fischer, \& Queloz}]{ufq07}
Udry, S., Fischer, D., \& Queloz, D. 2009, Protostars and Planets V (Tucson:
  Univ. Arizona Press)

\bibitem[{Udry \& Santos(2007)}]{us07}
Udry, S. \& Santos, N.~C. 2007, \araa, 45, 397

\bibitem[{Ward(1991)}]{ward91}
Ward, W.~R. 1991, in Luner and Planetary Institute Conference Abstract, 1463

\bibitem[{Ward(1997)}]{ward97}
---. 1997, Icarus, 126, 261

\end{thebibliography}

\clearpage

\begin{table}
\begin{center}
\caption{Summary of parameters \& symbols}
\label{param_disk}
\begin{tabular}{ccc}
\hline \hline
Symbol     & Meaning                        & Value                \\ \hline
$M_{* }$   & Stellar mass                   & 0.1 M$_{\odot }$       \\
$R_{*}$    & Stellar radius                 & 0.4R$_{\odot }$       \\
$T_{*}$    & Stellar effective temperature  & $2850$ K              \\
$\Sigma$   & The surface density of (gas + dust)     & $\propto r^{-1}$ \\
$M_d$    & The total disc mass (gas + dust) & $4.5\times 10^{-3}$ M$_{\odot}$   \\
           & gas-to-dust ratio              & 100                   \\
$\alpha$   & Parameter for turbulence (active/dead) & $10^{-2}/10^{-5}$ \\
$M_p$     & Planetary mass                 & 5 $M_{\oplus}$            \\
\hline
\end{tabular}
\medskip
\\
M$_{\odot }$ is one solar mass, and R$_{\odot }$ is one solar radius.  
\end{center}
\end{table}

\clearpage

\clearpage

\clearpage

\end{document}